\documentclass{AM}
\usepackage{float}
\usepackage{dcolumn,graphicx,color,booktabs,microtype,afterpage}
\usepackage{amssymb}
\usepackage{amsmath}
\usepackage[charter,greekuppercase=italicized]{mathdesign}
\usepackage[charter]{mathdesign}
\usepackage{sidecap}
\usepackage[mathlines]{lineno}
\usepackage[numbers,sort&compress]{natbib}
\usepackage{}
\graphicspath{{./}{figure/}}
\renewcommand{\tablename}{Table}
\makeatletter\renewcommand{\fnum@figure}[1]{\figurename~\thefigure.~}\makeatother
\makeatletter\renewcommand{\fnum@table}[1]{\tablename~\thetable.}\makeatother

\newcount\hh \newcount\mm
\hh=\time \divide\hh by 60
\mm=\hh \multiply\mm by 60 \mm=-\mm
\advance\mm by \time
\def\now{\number\hh:\ifnum\mm<10{}0\fi\number\mm}

\usepackage[colorlinks,plainpages=false,linkcolor=blue,urlcolor=blue,citecolor=blue,pdfpagemode=UseNone,pdfstartview=FitBH]{hyperref}

\newcommand{\tcr}[1]{\textcolor{black}{#1}}

\UseRawInputEncoding

\begin{document}
	
\pagestyle{fancy}
\rhead{\includegraphics[width=2.5cm]{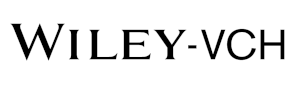}}	
\title{Engineering of orbital hybridization: an exotic strategy to \\ manipulate orbital current}
	
%
\maketitle

\author{Kun Zheng}
\author{Haonan Wang}
\author{Ju Chen}
\author{Hongxin Cui}
\author{Jing Meng}
\author{Zheng Li}
\author{Cuimei Cao}
\author{Haoyu Lin}
\author{Yuhao Wang}
\author{Keqi Xia}
\author{Jiahao Liu}
\author{Xiaoyu Feng}
\author{Hui Zhang}
\author{Bocheng Yu}
\author{Jiyuan Li}
\author{Yang Xu}
\author{Zhenzhong Yang*}
\author{Shijing Gong*}
\author{Qingfeng Zhan*}
\author{Tian Shang*}


%

\begin{affiliations}
	
K. Zheng, H. Cui, J. Meng, Z. Li, H. Lin, Y. Wang, K. Xia, J. Liu, H. Zhang, B. Yu, J. Li, Prof. Y. Xu, Prof. Q. Zhan,  Prof. T. Shang\\
Key Laboratory of Polar Materials and Devices (MOE), School of Physics and Electronic Science, East China Normal University, Shanghai 200241, China\\
Email Address: qfzhan@phy.ecnu.edu.cn, tshang@phy.ecnu.edu.cn

\vspace{1mm}
H. Wang, Prof. Z. Yang\\
Key Laboratory of Polar Materials and Devices (MOE), Shanghai Center of Brain-inspired Intelligent Materials and Devices, School of Physics and Electronic Science, East China Normal University, Shanghai 200241, China\\
Email Address: zzyang@phy.ecnu.edu.cn

\vspace{1mm}
J. Chen\\
School of Physics Science and Technology, Nantong University, Nantong 226019, China\\

\vspace{1mm}
C. Cao\\
Hubei Province Key Laboratory of Systems Science in Metallurgical Process, Wuhan University of Science and Technology, Wuhan 430081, China\\

\vspace{1mm}
X. Feng\\
State Key Laboratory for Artificial Microstructure \& Mesoscopic Physics and Frontiers Science Center for Nano-Optoelectronics, School of Physics, Peking University, Beijing 100871, China\\

\vspace{1mm}
Prof. S. Gong\\
Key Laboratory of Polar Materials and Devices (MOE), School of Physics and Electronic Science, East China Normal University, Shanghai 200241, China\\
Collaborative Innovation Center of Extreme Optics, Shanxi University, Taiyuan, Shanxi, 030006, China \\
Email Address: sjgong@ee.ecnu.edu.cn
\end{affiliations}



\begin{abstract}
\justifying
Current-induced spin-orbit torque (SOT) plays a crucial role in the next-generation spin-orbitronics. Enhancing its efficiency is both fundamentally and practically interesting, remains a challenge to date. Recently, orbital counterparts of spin effects which do not rely on the spin-orbit coupling (SOC) have been found as an alternative mechanism to realize it. This work highlights the engineering of copper oxidation states for manipulating the orbital current and its torque in the CuO$_x$-based heterostructures. The orbital hybridization and thus the orbital-Rashba-Edelstein effect at the CuO$_x$/Cu interfaces are significantly enhanced by increasing the copper oxidation state, yielding a torque efficiency that is almost 10 times larger than the conventional heavy metals. 
The Cu$_4$O$_3$/Cu interface rather than the widely accepted CuO/Cu interface is revealed to account for the enhanced SOT performance in the CuO$_x$-based heterostructures. In addition, the torque efficiency can be alternatively switched between high and low thresholds through the redox reaction. 
The current results establish an exotic and robust strategy for engineering the orbital current and SOT for spin-orbitronics, which applies to other weak-SOC materials.\\
\end{abstract}

\clearpage

\justifying
\section{Introduction}
Manipulating the magnetization via the interplay between spin, orbital, and charge degrees of freedom is one of the basics for spin-orbitronics~\cite{sinova_spin_2015,manchon_current-induced_2019,Shao2021}.
%
The current-induced spin-orbit torque (SOT) has been proposed as the key 
for the next-generation spin-orbitronics~\cite{miron_perpendicular_2011,liu_spin-torque_2012,tsai_electrical_2020,macneill_control_2017}, 
whose critical charge current density $J_c$ for switching the magnetization is two orders of magnitude smaller than that of conventional 
spin transfer torque~\cite{kiselev_microwave_2003,katine_current-driven_2000,myers_current-induced_1999}.
For a heterostructure consisting of heavy-metal (HM) and ferromagnetic (FM) layers, the charge current in the HM layer generates polarized spin current through either bulk or interfacial spin-orbit interaction (SOI), known as spin-Hall effect (SHE)\cite{hirsch_spin_1999,kato_observation_2004} or spin-Rashba-Edelstein effect (SREE) \cite{edelstein_spin_1990,mihai_miron_current-driven_2010}, which then exerts a torque (i.e., SOT) on the local magnetization in the FM layer (Figure~\ref{fig:DFT}a).

Enhancing the torque efficiency, which is both fundamentally and practically interesting, remains a challenge to date.
While materials with large spin-orbit coupling (SOC) are widely explored to achieve this goal~\cite{wang_field-free_2024,macneill_control_2017},
they face inevitable restrictions. For example, a larger SOC generally leads to a shorter spin-diffusion length~\cite{isasa_temperature_2015}. The electrical resistivity of those HMs is much larger than that of light elements, yielding a small spin-Hall angle, rarely exceeding 0.1~\cite{liu_current-induced_2012,wang_scaling_2014}. 
Such a strategy imposes a strong limitation on the material choices for spin-orbitronic devices, most of which contain 5$d$ heavy elements.

Recently, orbital counterparts of spin effects, namely orbital-Hall effect (OHE) and orbital-Rashba-Edelstein effect (OREE) that do not rely on the SOC, have been found as alternative mechanisms to overcome these constraints~\cite{Ando2025,Wang2025,choi_observation_2023,lee_orbital_2021,el_hamdi_observation_2023,ding_observation_2022}. Such orbital effects have even been argued as a fundamental mechanism of the spin effects~\cite{choi_observation_2023,el_hamdi_observation_2023}.
Orbital current, a flow of orbital angular momentum (OAM),
can be generated from charge current through either OHE or OREE. 
Such orbital current converts into spin current through the SOI in the SOC layer and is finally injected onto an FM layer and exerts an effective torque on the magnetization (Figure~\ref{fig:DFT}a), known as orbital torque (OT). Due to the giant orbital-Hall conductivity (OHC) of 3$d$ metals (e.g., Ti, Cr, Mn)\cite{go_first-principles_2024,salemi_first-principles_2022}, the orbital-driven mechanisms yield 
significantly enhanced torque efficiency (Figure~\ref{fig:DFT}b) and thus reduced current density $J_c$ (Figure~\ref{fig:DFT}c) than the spin-driven mechanisms. In addition, the light elements usually exhibit large orbital- and/or spin-diffusion length ($\sim$10-20 nm)~\cite{an_spintorque_2016,zheng_effective_2024}.

Interestingly, oxidation of light metals has been found to largely enhance SOT (e.g., open symbols in Figure~\ref{fig:DFT}b )~\cite{trier_oxide_2021,haku_role_2020,an_spintorque_2016,an_enhanced_2023},
%
highlighting the important role of oxides in spin-orbitronics. In particular, the CuO$_x$-based heterostructures represent one of the ideal platforms to explore the underlying physics of orbital effects at the interfaces in view of the relatively small OHC and thus weak OHE of element Cu~\cite{go_first-principles_2024,salemi_first-principles_2022}. 
Though different mechanisms have been proposed~\cite{kageyama_spin-orbit_2019,go_orbital_2021,okano_nonreciprocal_2019,gao_intrinsic_2018},
orbital current induced by the OREE at the CuO$_x$/Cu interface is believed to be the key for the enhanced torque~\cite{go_orbital_2021,ding_observation_2022}. 
Theoretical work reveals that such an OREE is closely related to the $p$--$d$ orbital hybridization at the CuO$_x$/Cu interface~\cite{go_orbital_2021}, 
which has not yet been experimentally explored. In previous studies, CuO$_x$ layer was introduced either by natural oxidation or by reactive sputtering in the mixed O$_2$/Ar atmosphere~\cite{an_spintorque_2016,gao_intrinsic_2018}, whose oxidation states are rather difficult to control. 
Though the naturally oxidized CuO$_x$ layer has been reported to significantly enhance the torque efficiency (Figure~\ref{fig:DFT}b)~\cite{an_spintorque_2016,ding_observation_2022,an_enhanced_2023,okano_nonreciprocal_2019},
the exact oxidation states or oxides that account for such an enhancement remain unexplored.  These two factors are crucial for exploiting the orbital transport and improving the SOT performance in spin-orbitronic devices. 

To clarify the above issues, here the copper oxidation states were systematically engineered through redox reaction, utilized by annealing the CuO$_x$-based heterostructures in the air and mixed Ar/H$_2$ atmospheres. 
Our results demonstrate that orbital hybridization and thus the OREE at the CuO$_x$/Cu interface are significantly enhanced by increasing the oxidation state of the CuO$_x$. As a consequence, the torque efficiency is largely enhanced. 
Furthermore, the Cu$_4$O$_3$/Cu interface, rather than the widely proposed CuO/Cu, was revealed to account for the enhanced SOT performance in the CuO$_x$-based heterostructures.
These findings not only validate the theoretical predictions regarding the orbital hybridization in enhancing the OREE and orbital current, but also establish an exotic and robust strategy toward high-efficiency spin-orbitronic devices based on manipulating the orbital degrees of freedom that are free of SOC.

\section{Results and discussion}
\begin{figure*}
	\centering 
	\includegraphics[width=0.95\linewidth]{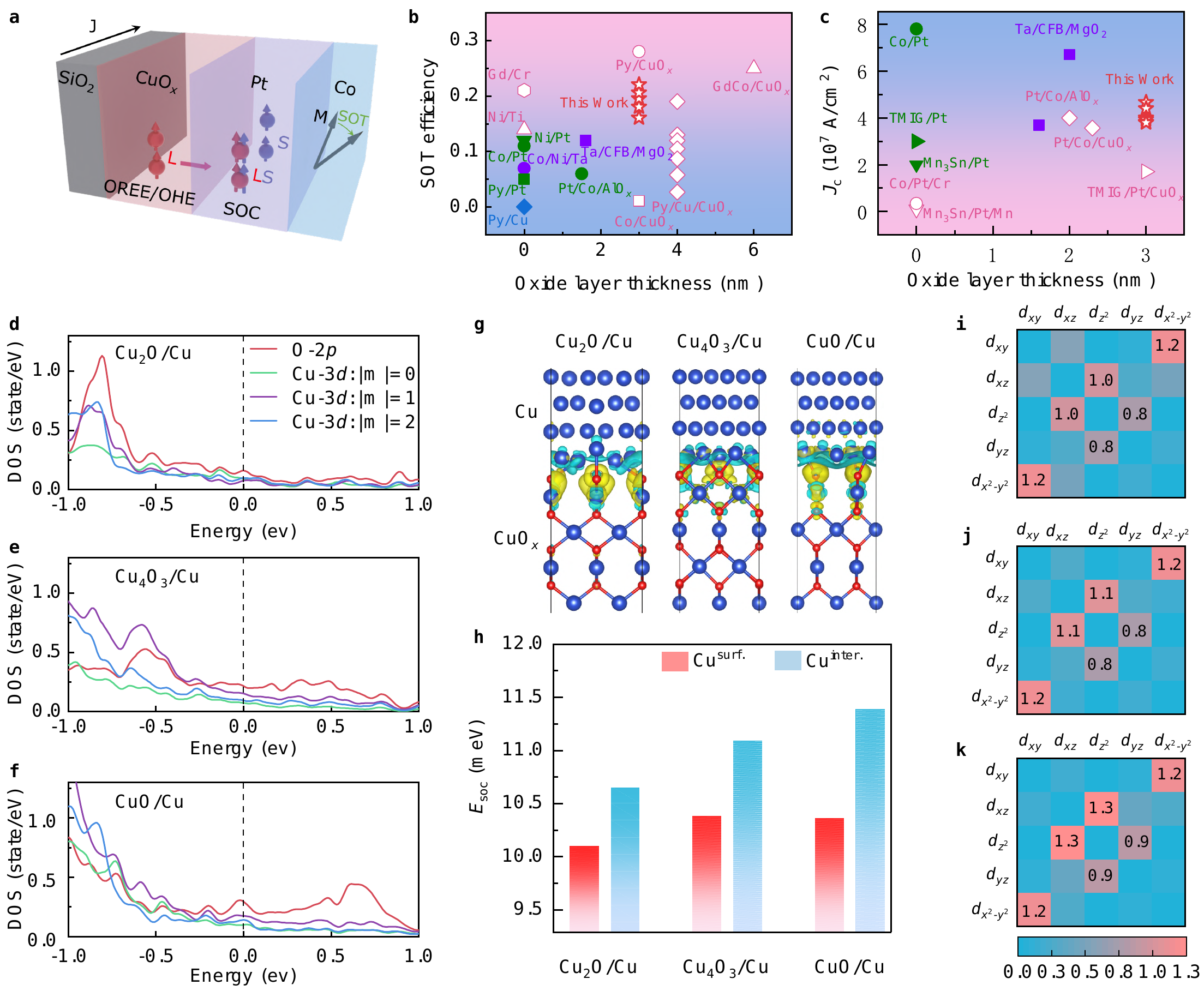}
	\caption{\label{fig:DFT} SOT properties of a variety of heterostructures and first-principles analysis of CuO$_x$/Cu interfaces. a) Schematic illustration of orbital-Hall effect or orbital-Rashba-Edelstein effect and orbital torque in Co/Pt/CuO$_x$ heterostructure. The orbital current is converted to a spin current through the SOC in Pt layer, and then exerts a torque on the Co moments. b,c) Torque efficiency (b) and critical current $J_c$ for magnetization switching versus the thickness of  an oxide layer for a variety of heterostructures. For torque efficiency in panel (b), solid symbols represent the heterostructures with heavy metals (e.g., Pt, Ta), while open symbols denote the heterostructures with a light-metal oxide layer (e.g., CuO$_x$). For $J_c$ in panel (c), solid and open symbols represent the heterostructures with dominant SHE and OHE (or OREE) mechanisms, respectively. 
		The star symbols represent the current work, while the data of other heterostructures were taken from Refs.~\cite{zhang_role_2015,cao_efficient_2022,ding_observation_2022,liu_spin-torque_2012,hu_efficient_2022,greening_current-induced_2020,an_enhanced_2023,an_electrical_2023,ding_orbital_2024,gao_intrinsic_2018,an_spintorque_2016,ding_unidirectional_2022,tsai_electrical_2020,zhang_electrical_2016,husain_field-free_2024,liu_current-induced_2012,ding_harnessing_2020,zheng_effective_2024,xiao_enhancement_2022}. The details are summarized in Table S1 in the Supporting Information.
		d-f) Density of states for the O-2$p$ and Cu-3$d$ orbitals at the interface of Cu$_2$O/Cu (d), Cu$_4$O$_3$/Cu (e), and CuO/Cu (f) heterostructures, respectively. Here, Cu$_2$O, Cu$_4$O$_3$, and CuO denote three different oxidation states of Cu ions. Their crystal structural information can be found in Table~S2 in the Supporting Information.
		g) Charge-density distributions at the Cu$_2$O/Cu, Cu$_4$O$_3$/Cu, and CuO/Cu interfaces. Yellow and blue regions indicate charge accumulation and depletion, respectively. 
		h) Spin-orbit coupling energy $E_\mathrm{SOC}$ for Cu$_2$O/Cu, Cu$_4$O$_3$/Cu, and CuO/Cu heterostructures. Blue and red bars represent the averaged $E_\mathrm{SOC}$ of Cu atoms at the CuO$_x$/Cu interfaces and at the vacuum-facing surfaces, respectively. 
		The $E_\mathrm{SOC}$ of each Cu atoms is listed in Table S3 in the Supporting Information.
		The optimized  CuO$_x$/Cu heterostructures are shown in Figure~S1 in the Supporting Information, and $E_\mathrm{SOC}$ of Cu$_2$O/Cu with two sets of different lattice constants are exceptionally discussed in Note~S1 in the Supporting Information.
		i-k) SOC matrix elements for different 3$d$ orbitals of Cu atoms in the Cu$_2$O/Cu (i), Cu$_4$O$_3$/Cu (j), and CuO/Cu (k) heterostructures.
	}
\end{figure*}

\noindent\textbf{2.1 Orbital hybridization at the CuO$_x$/Cu interface}
\vspace{3mm}\\
The previous theoretical work performed on the model of O/Cu(111) has demonstrated that the hybridization between O-$2p$ and Cu-$3d$ orbitals can be utilized to tune orbital current in metallic Cu film~\cite{go_orbital_2021}. 
To reveal the orbital hybridization at the CuO$_x$/Cu interface and its key role in enhancing the SOT, first-principles calculations were performed for Cu$_2$O/Cu, Cu$_4$O$_3$/Cu, and CuO/Cu heterostructures, characterized by different oxidation states in the CuO$_x$ layer. 
The projected density of states (DOS) of  the O-2$p$ and Cu-3$d$ orbitals with azimuthal angular momentum $|m|$ = 0 ($d_{z^2}$), 1 ($d_{xz}$ and $d_{yz}$), and 2 ($d_{x^2-y^2}$, $d_{xy}$) are shown in Figure~\ref{fig:DFT}d-f. Obviously, the O-2$p$ orbitals exhibit better extensibility in both Cu$_4$O$_3$/Cu and CuO/Cu heterostructures, which enhances the orbital hybridization. In particular,
in CuO/Cu with the highest copper oxidation state, the different orbitals show almost identical DOS profiles, facilitating a stronger $p$--$d$ hybridization. Figure~\ref{fig:DFT}g shows the charge density distribution at the various CuO$_x$/Cu interfaces. 
The electron depletion at the interfacial Cu (cyan) and accumulation at the O sites (yellow) are clearly distinct for the orbitals involved in the charge transfer among three different interfaces. 
In addition, variation of OAM can also be reflected by the SOC energy. 
The estimated SOC energy $E_\mathrm{SOC}$ of the interfacial metallic Cu atoms follows the sequence: 11.4~meV (CuO/Cu) $>$ 11.1~meV (Cu$_4$O$_3$/Cu) > 10.7~meV (Cu$_2$O/Cu) (blue bars in Figure~\ref{fig:DFT}h).
By contrast, the corresponding $E_\mathrm{SOC}$ of Cu atoms at the vacuum-facing surface is much smaller (red bars in Figure~\ref{fig:DFT}h). 
indicating that the $p$--$d$ hybridization clearly enhances the $E_\mathrm{SOC}$ of the interfacial Cu atoms. 
A more detailed orbital analysis of the interfacial Cu atoms (Figure~\ref{fig:DFT}i-k) suggest that the interactions between azimuthal angular momentum $|m|$ = 0 and $|m|$ = 1 play the pivotal role in determining $E_\mathrm{SOC}$ among three different CuO$_x$/Cu interfaces.
Though the $\left\langle d_{xy}|H_\mathrm{SOC}|d_{x^2-y^2}\right\rangle$ also shows comparable contributions,
it is almost invariable for three different CuO$_x$/Cu interfaces, and thus, it is not practically possible to account for the observed variation of OT in the CuO$_x$-based heterostructures. 
These calculations reveal the key role of $p$--$d$ orbital hybridization in determining the $E_\mathrm{SOC}$ and thus the orbital current. 

\vspace{3mm}
\noindent\textbf{2.2 Magnetization switching and SOT in naturally oxidized CuO$_x$-based heterostructures}
\vspace{3mm}\\
To experimentally verify the theoretical predictions and enhance the orbital current, the CuO$_x$ layer was introduced by natural oxidation of Cu, and its oxidation states were systematically controlled by annealing the heterostructures in the furnace (see details in the Methods and  Figure~S2, Supporting Information ).   
The time evolution of electrical resistance $R_\mathrm{xx}$ of the Pt/Co/Pt/CuO$_x$ (PCP/CuO$_x$) heterostructure in Figure~\ref{fig:Nat.Oxid}a
indicates that the CuO$_x$ layer prevents further oxidation of Cu after 48 hours of exposure in the air.
The perpendicular magnetic anisotropy (PMA) and the surface morphology of PCP/CuO$_x$ are robust against the natural oxidation (see inset in Figure~\ref{fig:Nat.Oxid}a and Figure~S3, Supporting Information). 
The PCP/CuO$_x$ exhibits comparable critical current density $J_c$ and switching ratio to the Pt/Co/Pt (PCP) heterostructure in the current-induced magnetization switching loop (Figure~\ref{fig:Nat.Oxid}b,c). Both heterostructures reach $\sim$80\% switching ratio in an assistance field of 7\,mT (Figure~\ref{fig:Nat.Oxid}d). 
 Moreover, field-free magnetization switching with a ratio of 4\% was observed in PCP/CuO$_x$, which is absent in the PCP heterostructure. 

To understand the field-free magnetization switching in PCP/CuO$_x$,
the spin-torque ferromagnetic resonance (ST-FMR) spectra were measured on Py/Pt/CuO$_x$ (Figure~\ref{fig:Nat.Oxid}e) and Py/Pt (Figure~\ref{fig:Nat.Oxid}f) heterostructures (see Figure~S5, Supporting Information). 
The determined torque efficiencies $\xi_\mathrm{FMR}$ are 0.16 and 0.08 for Py/Pt/CuO$_x$ and Py/Pt heterostructures, respectively.
Such an enhanced torque efficiency is attributed to the orbital current generated in the CuO$_x$ layer (Figure~\ref{fig:DFT}a), consistent with previous studies on the CuO$_x$-based heterostructures~\cite{ding_harnessing_2020,an_spintorque_2016}.
The opposite polarities of $R_\mathrm{xy}$--$J$ loop in Py/Pt/CuO$_x$ and PCP heterostructures is attributed to the dominant orbital current in the former case (Figure~\ref{fig:Nat.Oxid}b,c).
%
For Py/Pt, the ST-FMR spectra are almost identical when the external field is reversed (Figure~\ref{fig:Nat.Oxid}f), indicating that the torques are mostly attributed to the 
spin current with a $y$-axis spin polarization ($\sigma^y$).
By contrast, the Py/Pt/CuO$_x$ shows different spectra both in shape and amplitude for the opposite field directions (Figure~\ref{fig:Nat.Oxid}e), which are most likely ascribed to the $z$-axis spin polarization $\sigma^z$-induced torques~\cite{wang_field-free_2024,macneill_control_2017}. 
Such torques can be determined from  the measurements of angle-dependent ST-FMR spectra (Note~S2, Supporting Information). 
Though the $\sigma^z$-induced torques (dashed line) are much smaller than the $\sigma^y$-induced torques (dash-dotted line) (Figure~\ref{fig:Nat.Oxid}g), it nicely explains the field-free magnetization switching in Py/Pt/CuO$_x$ heterostructure. 
Such an out-of-plane contribution is negligible in the Py/Pt (Figure~S6, Supporting Information). 
The origins of spin current with $\sigma^z$ will be discussed below. 

\begin{figure}
	\includegraphics[width=0.95\linewidth]{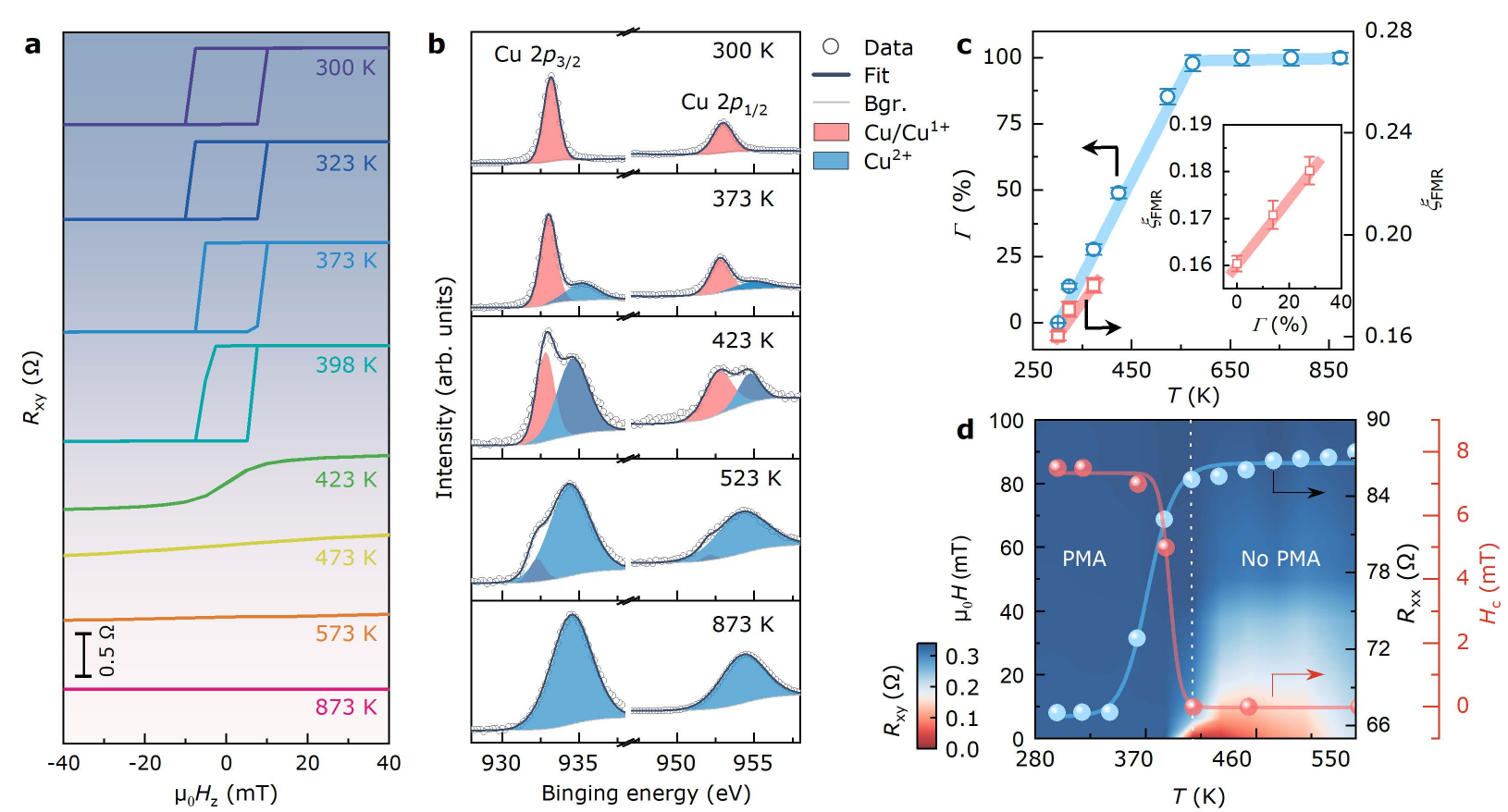}
	\caption{\label{fig:XPS} Characterization of perpendicular magnetic anisotropy and oxidation states.
		 a) Field-dependent Hall resistance $R_\mathrm{xy}(H)$ for 
		PCP/CuO$_x$ heterostructures annealed at different temperatures up to 873\,K in the air atmosphere. Except for 300\,K (i.e., natural oxidation), the PCP/CuO$_x$ heterostructures were annealed at each temperature for half an hour. For $R_\mathrm{xy}$ measurements, the magnetic field was applied along $z$-axis (i.e., perpendicular to the film plane). 
		b) X-ray photoelectron spectra of the Cu-2$p$ core-level of the PCP/CuO$_x$ heterostructures annealed at various temperatures. Solid black lines represent the fits including the contributions of Cu-2$p$ core-level of Cu$^{2+}$ (blue areas) and Cu/Cu$^{1+}$ (red areas). Multiple Gaussian distribution functions were used to fit the XPS spectra and to characterize the oxidation states of CuO$_x$ layer on the basis of Cu/Cu$^{1+}$ (933.5\,eV and 953.4\,eV) and Cu$^{2+}$ (934.4\,eV and 954.3\,eV) peaks. 
		The full XPS spectra, including the Cu$^{2+}$ satellite peaks, are presented in Figure~S8 in the Supporting Information.
		c) The peak-area ratio $\Gamma$ (left axis) 
		versus the annealing temperature for PCP/CuO$_x$ heterostructures.  
		The torque efficiency $\xi_{\mathrm{FMR}}$ of Py/Pt/CuO$_x$ heterostructures annealed at different temperatures are also summarized in panel (c) (right axis). Note that, as the annealing temperature increases above 450\,K, the contact electrodes (i.e., Ti/Cu) of SOT devices become fully oxidized and insulating, hindering further ST-FMR measurements. Inset shows the linear correlation between $\xi_{\mathrm{FMR}}$ and $\Gamma$ ratio, which suggests that $\xi_\mathrm{FMR}$ could be enhanced further if a proper electrode or FM layer is chosen. The ST-FMR spectra of these heterostructures are summarized in Figure~S9 in the Supporting Information.
		\tcr{d) Electrical resistance $R_\mathrm{xx}$ and coercive field $H_\mathrm{c}$ (right-axis) for the PCP/CuO$_x$ heterostructures as a function of annealing temperature.} The background color
		represents the magnitude of Hall resistance $R_\mathrm{xy}(H)$ of the PCP/CuO$_x$ heterostructures annealed at various temperatures. Here, the $R_\mathrm{xy}$ data collected from 100 to 0\,mT are solely presented. 
		Note the significant suppression of spontaneous anomalous Hall resistance and thus of the PMA  in PCP/CuO$_x$ heterostructures when the annealing temperature is higher than 423\,K, as indicated by the dashed line. All the XPS, ST-FMR, $R_\mathrm{xx}$ and $R_\mathrm{xy}$ measurements were carried out at room temperature. 
	}
\end{figure}

\vspace{3mm}
\noindent\textbf{2.3 Manipulation of oxidation states in the CuO$_x$ layer}
\vspace{3mm}\\
To manipulate the oxidation state of the CuO$_x$ layer, after 24 hours of natural oxidation, the PCP/CuO$_x$ heterostructures were annealed at various temperatures up to 873\,K in the air atmosphere. The PMA of PCP layer is robust against the annealing temperature at $T <$ 423\,K. However, it is destroyed when the annealing temperature is higher than 423\,K, reflected by the absence of spontaneous anomalous Hall resistance $R_\mathrm{xy}^\mathrm{A}$ in PCP/CuO$_x$ (Figure~\ref{fig:XPS}a,d). 
For the high-temperature annealing, 
the Co layer ($\sim$0.7\,nm) could be oxidized and become antiferromagnetic~\cite{tomiyasu_magnetic_2004}, and thus, only the ordinary Hall resistance was detected (Figure~S7, Supporting Information). 
The $R_\mathrm{xx}$ undergoes a step-like increase as the annealing temperature exceeds 350\,K and starts to saturate at $T \ge$ 423\,K, where the spontaneous $R_\mathrm{xy}^\mathrm{A}$ disappears (Figure~\ref{fig:XPS}d). 

The oxidation states of the CuO$_x$ layer were characterized by the x-ray photoelectron spectroscopy (XPS) measurements.
For the naturally oxidized PCP/CuO$_x$ heterostructure, the XPS spectra show two distinct peaks at binding energies of 933.5 and 953.4\,eV (Figure~\ref{fig:XPS}b), which correspond to the core level of Cu-$2p_{3/2}$ and Cu-$2p_{1/2}$, respectively~\cite{an_enhanced_2023,maack_oxidation_2018}. As the annealing temperature increases, both single peaks split into double peaks. For 873\,K, the CuO$_x$ layer is fully oxidized, the single peaks are restored but shift to higher binding energies at 934.4 and 954.3\,eV. 
\tcr{It is noted that Cu$^{1+}$ peak is rather difficult to distinguish from the Cu peak~\cite{an_enhanced_2023,maack_oxidation_2018,zhang_electronic_2025,ramirez_oxidation_2011}.} However, the relative Cu$^{2+}$ oxidation state in the CuO$_x$ layer can be quantified by a peak-area ratio $\Gamma$ = $S_\mathrm{Cu^{2+}}$/($S_\mathrm{Cu/Cu^{1+}}$+$S_\mathrm{Cu^{2+}}$), where $S_\mathrm{Cu^{2+}}$ and $S_\mathrm{Cu/Cu^{1+}}$ represent the total areas of Cu$^{2+}$ (blue) and Cu/Cu$^{1+}$ (red) peaks in the XPS spectra (Figure~\ref{fig:XPS}b). 
The $\Gamma$ ratio linearly increases with the annealing temperature and reaches almost 100\% when the annealing temperature exceeds 573\,K (Figure~\ref{fig:XPS}c). 
These results confirm that all the CuO$_x$ layer reaches the highest oxidation state (i.e., CuO) when the annealing temperature is larger than 573\,K, consistent with the $R_\mathrm{xx}$ results in Figure~\ref{fig:XPS}d. 
Interestingly, $\xi_\mathrm{FMR}$ shows linear annealing-temperature or $\Gamma$-ratio dependence, reaching $\sim$0.18 in the Py/Pt/CuO$_x$ heterostructure annealed at 373\,K (Figure~\ref{fig:XPS}c). Such $\Gamma$-dependent $\xi_\mathrm{FMR}$ implies an intimate relationship between orbital torque and oxidation state of the CuO$_x$ layer.

\vspace{3mm}
\noindent\textbf{2.4 Orbital torque tuned by orbital hybridization}
\vspace{3mm}\\
Based on the above results, to maintain the PMA and to avoid the oxidation of contact electrodes and of magnetic layer, 
the annealing temperature was fixed at 373\,K but the annealing time was prolonged to optimize the SOT performance for the CuO$_x$-based heterostructures.  
The PMA and magnetization of PCP/CuO$_x$ heterostructures are robust against the annealing time (Figure~S10, Supporting Information). 
The determined $\xi_\mathrm{FMR}$ of Py/Pt/CuO$_x$ increases with annealing time (Figure~\ref{fig:orbit_torque}a and Figures~S11 and S12, Supporting Information), resulting in a decreasing (increasing) in $J_c$ (switching ratio) in the field-free magnetization switching for PCP/CuO$_x$ (Figure~\ref{fig:orbit_torque}b,c).
Due to the enhanced torque efficiency, the SOT performance is significantly improved for those CuO$_x$-based heterostructures.
For example, after annealing the heterostructures at 373\,K for 2.5 hours, the switching ratio ($\sim$18\%) is increased by a factor of 350\%, while the $J_c$ ($\sim$3.79$\times$10$^7$~Acm$^{-2}$) is reduced by a factor of 19\% compared to the naturally oxidized heterostructures. The switching ratio also reaches 100\% in an assistance field of 10\,mT for these annealed heterostructures (Figure~S13, Supporting Information).

Following the first-principles calculations (Figure~\ref{fig:DFT}) and XPS measurements (Figure~\ref{fig:XPS}), 
the $\Gamma$ ratio can be regarded as a scale of orbital hybridization at the CuO$_x$/Cu interfaces. 
The $\Gamma$ ratio of PCP/CuO$_x$ increases as prolonging the annealing time and reaches $\sim$50\% after 2.5-hours annealing (Figure~S15, Supporting Information). Therefore, the $\xi_\mathrm{FMR}$ increases with the $\Gamma$ ratio (Figure~\ref{fig:orbit_torque}d), resembling the results of CuO$_x$-based heterostructures annealed at different temperatures (Figure~\ref{fig:XPS}c). 
The spin-Hall angle $\theta_\mathrm{SH}$ determined by the harmonic-Hall measurements in PCP/CuO$_x$ heterostructures (Figures~S16 and S17, Supporting Information) shows similar $\Gamma$ dependence as the $\xi_\mathrm{FMR}$ of Py/Pt/CuO$_x$ (Figure~\ref{fig:orbit_torque}d).
The slightly reduced torque efficiency in PCP/CuO$_x$ heterostructures is attributed to the cancellation of spin current in the bottom and top Pt layers. The torque efficiencies of CuO$_x$-based heterostructures are significantly enhanced compared to the conventional HM-based heterostructures (see Figure~\ref{fig:DFT}b).
The in-plane ($\xi_\mathrm{DL}^y$) and out-of-plane ($\xi_\mathrm{DL}^z$) damping-like torques of Py/Pt/CuO$_x$ were extracted by the angular-dependent ST-FMR measurements (Figure~S15 and Note~2, Supporting Information). Though the $\xi_\mathrm{DL}^z$ is almost 
two orders of magnitude smaller than the $\xi_\mathrm{DL}^y$, both of which exhibit similar $\Gamma$ dependence (Figure~\ref{fig:orbit_torque}e). The enhanced $\xi_\mathrm{DL}^z$ nicely explains the results of field-free magnetization switching in PCP/CuO$_x$ heterostructures (Figure~\ref{fig:orbit_torque}c). 


The orbital torque in CuO$_x$-based heterostructures can be systematically controlled by manipulating the oxidation states of CuO$_x$ layer through the redox reaction. The oxidation states of CuO$_x$ layer can be enhanced or reduced by annealing the heterostructures in the air or mixed Ar/H$_2$ atmosphere (see details in Figures~S19 and 20, Supporting Information). Interestingly, 
the high- and low-torque states can be achieved through the redox-reaction cycles (Figure~\ref{fig:orbit_torque}f). Such an "orbital torque" switch is rather robust, with the reproducibility maintained for the cycle number up to 5.   
Moreover, similar results were obtained in the magnetization switching ratio (Figure~S21, Supporting Information). 
Combined with theoretical calculations, the above results provide solid evidence that the orbital hybridization at the Cu/CuO$_x$ interfaces plays a crucial role in generating the orbital current in the CuO$_x$-based heterostructures. These findings offer an avenue for engineering and reversible manipulation of the orbital current and SOT for spin-orbitronic devices.

\begin{figure}
		\centering
	\includegraphics[width=0.8\linewidth]{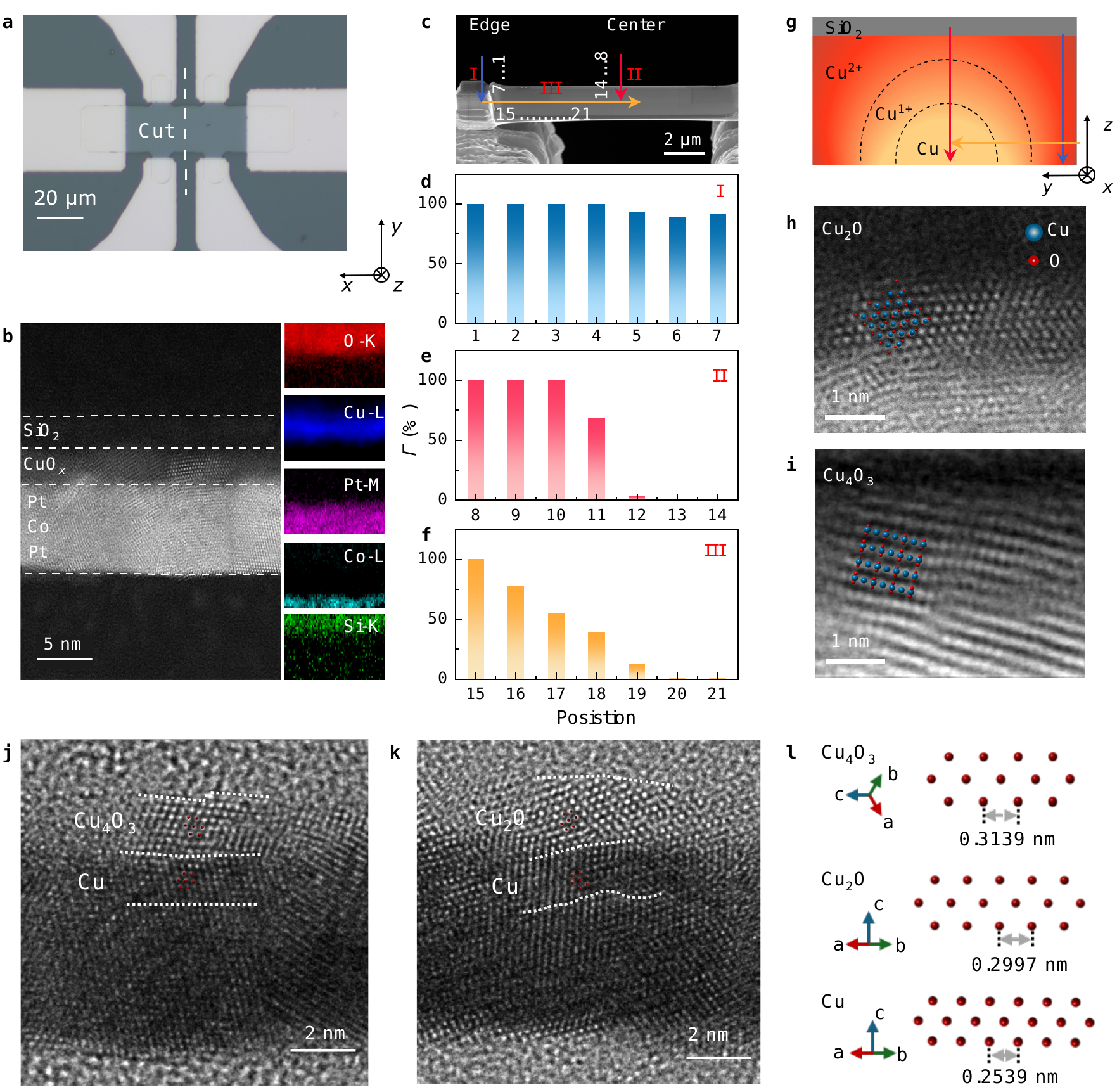}
	\caption{\label{fig:TEM} Evidence of different oxidation states and their distributions in the CuO$_x$ layer.
		%
		 a) A representative optical image of the micro-fabricated PCP/CuO$_x$ heterostructure with a Hall-bar geometry, which was
		annealed at 373\,K for 2 hours and was used for the TEM measurements. The dashed line indicates the cut direction by using a focused ion beam.  The determined $\Gamma$ ratio is close to 50\% (Figure~S15, Supporting Information) for this device. 
		b) Cross-sectional HAADF-STEM image with EELS elemental maps. The distributions of different elements including O (red), Cu (blue), Pt (purple), Co (cyan), and Si (green) are present. 
		The fast Fourier transform images of PCP and CuO$_x$ layers are presented in Figure~S22, Supporting Information.
		c) Cross-sectional SEM image. The arrows mark the scanning paths (I, II, and III) of the EELS measurements. 
		Paths I (points 1-7) and II (points 8-14) scan from SiO$_2$/CuO$_x$ to CuO$_x$/Pt interfaces near the edge and in the center of the cut PCP/CuO$_x$ heterostructure, while the path III (points 15-21) scans from the edge to the center within the CuO$_x$ layer.  
		d-f) The $\Gamma$ ratios (see definition in Figure~\ref{fig:XPS}) as determined from EELS spectra collected along paths I (d), II (e), and III (f) at different positions marked in panel (c).
		Path I (blue arrow) and path II (red arrow) are close to the edge and center of the Hall bar, respectively, and EELS spectra were collected from the top to the bottom of CuO$_x$ layer; Path III is near the CuO$_x$/Pt interface, and EELS spectra were collected from the edge to the center of the Hall bar. 
		Similar to the XPS spectra (Figure~\ref{fig:XPS}b), the peaks at energies of 933.5\,eV and 934.4\,eV in the EELS spectra are attributed to the Cu $L$-edge for Cu/Cu$^{1+}$ and Cu$^{2+}$ oxidation states, respectively.
		The EELS spectra fitted with multiple Gaussian distribution functions are summarized in Figure~S23, Supporting Information. 
		%
		g) Schematic plot of the CuO$_x$ layer demonstrating the distributions of Cu$^{2+}$, Cu$^{1+}$, and Cu. The arrows mark the same paths as in panel (c). 
		h,i) Atomic-resolution HAADF-STEM images and the corresponding crystal models
		for Cu$_2$O (h) and Cu$_4$O$_3$ (i), which are oriented along [110] and [011] directions, respectively. Blue and red spheres represent Cu and O atoms, respectively. 
		The Cu$_2$O and Cu$_4$O$_3$ crystallize in cubic ($Pn3m$) and  tetragonal ($I4_1$/$amd$) crystal structures (see crystallographic information in Supporting Information Table~S2).
		No CuO nanoparticles could be identified in the CuO$_x$ layer, consistent with x-ray diffraction measurements (see Figures~S25 and S26, Supporting Information). \tcr{j,k) HR-TEM image of the PCP/CuO$_x$ heterostructure annealed at 373 K for 2 hours. Both Cu$_4$O$_3$/Cu and Cu$_2$O/Cu interfaces can be clearly tracked. l) Crystal structures of Cu$_4$O$_3$ (top) viewed along the [111], Cu$_2$O (middle) viewed along the [110], and Cu (bottom) viewed along the [110] zone axis, respectively. Here, only the Cu atoms are presented. The crystal structures and atomic distances match the HR-TEM image very well, confirming the presence of both Cu$_4$O$_3$/Cu and Cu$_2$O/Cu interfaces.}
		}	
\end{figure}

\vspace{3mm}
\noindent\textbf{2.5 Microscopic evidence of oxidation states and their distributions in the CuO$_x$ layer}
\vspace{3mm}\\
%
%
 %
Both XPS and ST-FMR measurements reveal a strong correlation between Cu$^{2+}$ oxidation state and torque efficiency in the CuO$_x$-based heterostructures. The theoretical calculations also suggest that the $p$--$d$ hybridization (Figure~\ref{fig:DFT}d-f) and $E_\mathrm{soc}$ of metallic Cu (Figure~\ref{fig:DFT}h) are most significant for the highest oxidation state. 
%
However, the XPS spectra fail to distinguish the CuO from Cu$_4$O$_3$ (Figure~\ref{fig:XPS}b), the latter can be regarded as a mixture of Cu$^{2+}$ and Cu$^{1+}$ states. 
The cross-sectional transmission electron microscopy (TEM) measurements were performed on PCP/CuO$_x$ device (Figure~\ref{fig:TEM}a) to identify which CuO$_x$/Cu interface that enhances the orbital torque. 
Cross-sectional high-angle annular dark field-scanning transmission electron microscopy (HAADF-STEM) images reveal that both the PCP and CuO$_x$ layers are crystallized after the annealing (Figure~\ref{fig:TEM}b and Figure~S22, Supporting Information), contrasting with the amorphous structure in the naturally oxidized CuO$_x$-based heterostructures~\cite{xiao_enhancement_2022}. 

The electron energy-loss spectroscopy (EELS) mapping reveals 
the distributions of different atoms in each layer (Figure~\ref{fig:TEM}b). The EELS spectra were also collected along different paths indicated by the arrows in Figure~\ref{fig:TEM}c. 
The estimated $\Gamma$ ratios from EELS spectra (Figure~S23, Supporting Information) show distinct variations at different positions along three paths. For path I, the $\Gamma$ ratio is close to 100\%, and is almost independent of positions (Figure~\ref{fig:TEM}d); For path II, close to the CuO$_x$/SiO$_2$ interface, the $\Gamma$ ratio is also close to  100\%. However, it significantly decreases and becomes zero when moving to the CuO$_x$/Pt interface (Figure~\ref{fig:TEM}e); For path III, the $\Gamma$ ratio 
continuously decreases from 100\% to 0\% when moving from the edge to center of the Hall bar (Figure~\ref{fig:TEM}f).     %
Based on the  EELS spectra, the dome-like distributions of different copper oxidation states were constructed (Figure~\ref{fig:TEM}g). 
\tcr{Although the dome-like oxidation gradient makes the determination of the CuO$_x$ layer thickness challenging, it can be estimated by using EELS mapping and line profile analysis (Figure~S24, Supporting Information), which is about 1.3\,nm in the PCP/CuO$_x$ heterostructure annealed at 373 K for 2 hours.}


Within the CuO$_x$ layer, the single-crystalline grains of Cu$_2$O (Figure~\ref{fig:TEM}h) and Cu$_4$O$_3$ (Figure~\ref{fig:TEM}i) can be clearly identified in the atomic-resolution HAADF-STEM images, while no CuO grains can be spotted. 
\tcr{Furthermore, high-resolution TEM (HR-TEM) imaging reveals the Cu$_4$O$_3$/Cu and Cu$_2$O/Cu interfaces within the CuO$_x$ layer (see Figure~\ref{fig:TEM}j-k).}
The absence of CuO is further confirmed by x-ray diffraction measurements (Figures~S25 and S26, Supporting Information). The CuO reflections start to appear when the annealing temperature is above 423\,K, and a Cu film is fully oxidized into a CuO film as the annealing temperature is higher than 623\,K, consistent with the XPS results in Figure~\ref{fig:XPS}b.
Therefore, the observed Cu$^{2+}$ oxidation state in the CuO$_x$-based heterostructures annealed at temperatures lower than 423\,K is attributed to the Cu$_4$O$_3$. 
 This is also the case for the naturally oxidized CuO$_x$, where CuO has been widely accepted to account for the enhanced torque in the previous studies~\cite{an_enhanced_2023,gao_intrinsic_2018,kageyama_spin-orbit_2019}. 
These TEM and XRD results clearly clarify that the Cu$_4$O$_3$/Cu interface plays a crucial role in enhancing the orbital current in the CuO$_x$-based heterostructure, a key factor that has never been clarified in the past.
\section{Discussion}
Several approaches have been applied to enhance the OT or SOT, those including the utilizing of light elements with a giant OHC (e.g., Cr, Mn)~\cite{go_first-principles_2024,salemi_first-principles_2022} and FM materials with a large orbit-spin conversion $\eta_\mathrm{LS}$ (e.g., Ni, Gd)~\cite{lee_orbital_2021,lee_efficient_2021}, as well as enhancing the orbital hybridization at the interfaces~\cite{an_spintorque_2016,ding_harnessing_2020}.
Our first-principles calculations demonstrate that the orbital hybridization between Cu-3$d$ and O-2$p$ orbits is clearly distinct for different CuO$_x$/Cu interfaces. 
The $E_\mathrm{SOC}$ of the metallic Cu layer and the $p$--$d$ hybridization increase as the oxidation state of the CuO$_x$ is enhanced. Experimentally, the oxidation state of the CuO$_x$ can be manipulated through redox reaction, yielding tunable SOT performance. After systematically optimizing the annealing conditions, the torque efficiency is enhanced to $\xi_\mathrm{FMR}$ $\sim$ 0.22 for the Py/Pt/CuO$_x$ heterostructure, which is almost 4 times larger than the conventional HM-based heterostructures~\cite{liu_spin-torque_2012,liu_current-induced_2012}, and almost an order of magnitude larger than that of the Pt(1.5)/Py(11) bilayer ($\xi_\mathrm{FMR}$ $\sim$ 0.03). 
\tcr{According to the XRD measurements (Figures S25 and S26, Supporting Information), the lattice constants of the annealed Cu films are comparable to the value of bulk Cu metal and are almost independent of annealing temperature and time. Therefore, the annealing-induced strain/stress effects or structural distortions can be excluded in the CuO$_x$-based heterostrcuture.}
\tcr{The high-field Hall-resistivity measurements (Figure S27, Supporting Information) yield an almost constant carrier density for the PCP/CuO$_x$ heterostructures. The changes in the electrical conductivity or the carrier density barely account for the enhanced torque efficiency.}
The XPS measurements reveal that the fraction of Cu$_4$O$_3$ reaches $\sim$50\% in the CuO$_x$ layer in such a heterostructure. Considering that both spin- and orbital Hall conductivities of Cu are significantly smaller than Pt~\cite{salemi_first-principles_2022}, its spin- and orbital Hall effects barely account for the largely enhanced SOT in the CuO$_x$-based heterostructures. 	    


Since the SOC of Cu is negligible, the SREE at the CuO$_x$/Cu interface can be excluded. The SOT caused by the SREE at the Cu/Pt interface should be much smaller than the bulk SHE-induced SOT in the Pt layer~\cite{liu_current-induced_2012,salemi_first-principles_2022}.
As an alternative mechanism, the OREE at the CuO$_x$/Cu interface is most likely the origin for the significantly enhanced torques. 
The OREE is closely related to the orbital hybridization at the CuO$_x$/Cu interface~\cite{go_orbital_2021}, which can be manipulated by the oxidation state in the CuO$_x$ layer.  
As a consequence, the OREE and thus the orbital current can be systematically controlled through redox reaction, 
resulting in a switchable torque efficiency (Figure~\ref{fig:orbit_torque}f).
The oxidation state of the CuO$_x$ layer can also be enhanced either by increasing the O$_2$ flow rate during the reactive sputtering or by increasing the ionic-liquid gating voltage in the Py/CuO$_x$ heterostructures~\cite{an_electrical_2023,kageyama_spin-orbit_2019,gao_intrinsic_2018}. 
However, the CuO$_x$ layer deposited by reactive sputtering is quite insulating, yielding small and almost $\Gamma$-independent torque efficiencies (see grey and green crosses in Figure~\ref{fig:orbit_torque}d,e)~\cite{kageyama_spin-orbit_2019,gao_intrinsic_2018}.
Besides, the CuO$_x$ layer deposited by reactive sputtering is homogeneous, and thus an assistance field is required to switch the magnetization.
For the ionic-liquid gating, the $\Gamma$ ratio increases with the gating voltage~\cite{an_electrical_2023}, leading to an enhanced $\xi_\mathrm{FMR}$ $\sim$ 0.1 too (see yellow crosses in Figure~\ref{fig:orbit_torque}d). 
However, such a gating technique shows low repeatability, which is unsuitable for the spin-orbitronic applications~\cite{petach_disorder_2017}.
Moreover, the intrinsic Berry curvature mechanism has also been proposed to explain the SOT in the CuO$_x$-based heterostructures~\cite{gao_intrinsic_2018}.
Our results offer a new pathway 
for engineering and reversible manipulation of the orbital current and SOT for high-efficiency spin-orbitronic devices.

The enhanced OREE through orbital hybridization at the CuO$_x$/Cu interfaces is further supported by ST-FMR measurements on CuO/Pt/Py, Cu$_2$O/Cu/Pt/Py, and CuO/Cu/Pt/Py heterostructures, where the CuO and Cu$_2$O layers were directly deposited using the targets. 
Both Cu$_2$O and CuO films are very insulating, therefore, both CuO/Pt/Py and Pt/Py heterostructure show comparable $\xi_\mathrm{FMR}$~\cite{liu_current-induced_2012}. For CuO/Cu/Pt/Py, the $\xi_\mathrm{FMR}$ $\sim$ 0.14 is twice larger than that of 
Cu$_2$O/Cu/Pt/Py heterostructure (Figure~S28, Supporting Information), consistent with the most significant $p$--$d$ orbital hybridization and the OREE at CuO/Cu interface (Figure~\ref{fig:DFT}h). 
\tcr{In addition, both the Cu/Py ($\xi_\mathrm{FMR}$ $<$ 0.001) and the Pt/Py (or Cu/Pt/Py) ($\xi_\mathrm{FMR}$ $\sim$ 0.05) heterostructures exhibit significantly smaller torque efficiency, where the SHE is expected to be the driven mechanism.}
It is noted that the CuO-based heterostructure cannot be achieved by annealing procedures due to oxidation of  either contact electrodes or the magnetic layers (i.e., Py and Co).     

%
%
%

Finally, we discuss the field-free magnetization switching in the PCP/CuO$_x$ heterostructures. Such a magnetic switching requires the out-of-plane spin polarization $\sigma^z$, generally induced by breaking either the time-reversal or the inversion symmetry~\cite{yu_switching_2014,wang_field-free_2024}. 
Since the magnetic ordering temperatures of CuO$_x$ are well below the room temperature~\cite{Djurek2015}, the magnetic origins could be excluded.
In addition to the intrinsic noncentrosymmetric materials (e.g., WTe$_2$~\cite{wang_field-free_2024,macneill_control_2017}), 
the inversion symmetry also can be broken in the presence of composition gradient~\cite{yu_switching_2014} or electric field on the surface of oxide substrates~\cite{wang_field-free_2023}.	
In the case of CuO$_x$-based heterostructures, the systematic annealing creates a dome-like oxidation gradient in the CuO$_x$ layer (Figure~\ref{fig:TEM}g).
Such an oxidation gradient breaks the inversion symmetry in the CuO$_x$ layer, resulting in a $\sigma^z$ that is absent in the naturally oxidized CuO$_x$-based heterostructures~\cite{ding_harnessing_2020,an_spintorque_2016}. 
This is further supported by the absence of $\sigma^z$ in the CuO/Cu/Pt/Py and Cu$_2$O/Cu/Pt/Py heterostructures, where the homogeneous CuO and Cu$_2$O layers were deposited using the targets (Figure~S29, Supporting Information).
As an alternative, spin-vorticity coupling (SVC) mechanism has been proposed to explain the SOT in CuO$_x$-based heterostructures~\cite{okano_nonreciprocal_2019,an_enhanced_2023}. 
The oxidation gradient in the CuO$_x$ layer leads to the formation of charge current vorticity, which in turn couples to the spin degree of freedom and generates the spin current. A dome-like oxidation gradient allows the charge current vorticity along both $y$- and $z$-axes, resulting in spin current with both  $\sigma^y$  and $\sigma^z$, the latter accounting for the field-free magnetization switching.
Considering that all the copper oxides are insulating, comparable electrical conductivity gradients are expected for the Cu$_2$O/Cu, Cu$_4$O$_3$/Cu, and CuO/Cu interfaces. Therefore, the largely enhanced torque efficiency in the CuO$_x$ based heterostructures is mainly attributed to the enhanced OREE rather than the SVC mechanism.
In addition, the Cu-based heterostructures that were annealed in air show significantly smaller torque efficiency $\xi_\mathrm{FMR}$ $\sim$ 0.07 than the CuO$_x$-based heterostructures (see  Figures~S30-S32, Supporting Information), highlighting the crucial role for introducing oxygen in the CuO$_x$ layer by natural oxidation. \tcr{In addition, the possible Joule heating effects are evaluated in the current-induced magnetization switching measurements. The maximum temperature rise due to Joule heating is $\sim$50-60 K for all the PCP/CuO$_x$ heterostructures (see details in Note S3 and Table S5, Supporting Information). According to the Hall-resistivity measurements at $T$ = 300 and 350\,K, both the remanent Hall resistivity and coercive field slightly decrease with temperature. However, for a fixed temperature, the  remanent Hall resistivity and coercive field are almost constant for all the PCP/CuO$_x$ heterostructures (Figure~S33, Supporting Information). Considering that comparable thermal effects are expected in all the PCP/CuO$_x$ heterostructures during the  magnetization switching measurements, the reduced critical current density is most likely attributed to the enhanced torque efficiency.
}

\section{Conclusion}


To summarize, our work demonstrates that the efficiency of charge-to-orbital current conversion can be systematically tuned by engineering the copper oxidation states in the CuO$_x$-based heterostructures. This conversion is determined by the orbital hybridization and the OREE at CuO$_x$/Cu interfaces, both of which are significantly enhanced with copper oxidation states. Consequently, OT and/or SOT is largely enhanced, leading to a reduced $J_c$ for current-induced magnetization switching. TEM measurements reveal that the Cu$_4$O$_3$/Cu interface rather than the widely accepted CuO/Cu accounts for the enhanced SOT performance in the CuO$_x$-based heterostructures. Furthermore, the torque efficiency can be alternatively switched between high and low thresholds through the redox reaction. These findings establish an easy but robust strategy for engineering and reversibly manipulating orbital current and torque for spin-orbitronic applications. The nonmagnetic 3$d$ metals, such as Sc, V, and Ti, whose OHC is significantly larger than that of Cu, represent one of the ideal candidates to apply such a strategy and to further optimize the SOT performance.

\section{Experimental Section}
\noindent\threesubsection{Thin-film deposition and device fabrication}\\
Heterostructures of Pt(3)/Co(0.7)/Pt(1.5)/CuO$_x$(3), Py(11)/Pt(1.5)/CuO$_x$(3), 
Py(11)/Pt(1.5)/Cu/Cu$_2$O(3), Py(11)/Pt(1.5)/Cu/CuO(3), Py(11)/Pt(1.5), and Pt(3)/Co(0.7)/Pt(1.5) (bracketed numbers denote thicknesses in nm)  were deposited on thermally oxidized Si substrate at room temperature in an ultra-high vacuum magnetron sputtering system with a base pressure below $5.0 \times 10^{-8}$~\,Torr. The Pt/Co/Pt/CuO$_x$ heterostructures were used to investigate the current-induced magnetization switching, while the Py/Pt/CuO$_x$ heterostructures were fabricated to evaluate the torque efficiency. A 2-nm-thick SiO$_2$ capping layer was deposited on top of these heterostructures. 
The CuO$_x$ layer was produced through natural oxidation of the Cu layer in air before the deposition of SiO$_2$ capping layer. 
To increase (or decrease) the oxidation state of the CuO$_x$ layer, the heterostructures were annealed in the air [or mixed (95\%)/H$_2$(5\%)] atmosphere. The surface morphology of the deposited heterostructures were checked by atomic force microscopy (Bruker, Multimode 8) measurements. The XRD patterns were collected using a PANalytical X'Pert Pro x-ray diffractometer with Cu K$\alpha$ radiation ($\lambda$ = 1.5418~\AA{}). The XPS measurements were performed using a Thermo Scientific ESCALAB Xi+ system equipped with a monochromatic Al K$\alpha$ x-ray source (h$\nu$ = 1486.6~eV).

To perform the transport measurements, the heterostructures were patterned into a Hall-bar configuration (central area: 20 {\textmu}m $\times$ 100 {\textmu}m; electrodes: 150 {\textmu}m $\times$ 150 {\textmu}m) (see Figure~S2, Supporting Information) using 
photolithography and Ar ion etching techniques. While for the 
torque measurements, the heterostructures were patterned into a rectangle bar with dimensions of 20 {\textmu}m $\times$ 100 {\textmu}m (see Figure~S5, Supporting Information). The contact electrodes consisted of a 10 nm Ti layer capped with a 100 nm Cu layer. 
\\

\noindent\threesubsection{Magnetization measurements}\\
Field-dependent magnetizations $M(H)$ data were collected at room temperature by applying in-plane magnetic fields using a Quantum Design magnetic property measurement system. The heterostructures with typical dimensions of 5 $\times$ 5\,mm$^2$ were used for the magnetization measurements.\\



\noindent\threesubsection{Current-induced magnetization switching measurements}\\
For current-induced magnetization switching measurements, a Keithley 6221 source meter was used as the current source for applying both the DC and the AC electric currents. A series of current pulses ($J$) with a 100~{\textmu}s pulse was applied to switching the magnetization, while a DC current of 2~mA was applied to characterize the magnetization switching through the measurements of anomalous Hall resistance. The Hall voltage  was collected using a Keithley 2182A nanovoltmeter. The assistance field ($H_x$) was applied parallel to the current direction. \\

\noindent\threesubsection{ST-FMR measurements}\\
The ST-FMR spectra were measured by a lock-in amplifier. A microwave-frequency (RF)
charge current $I_\mathrm{rf}$ was applied along the longitudinal direction of the device.
The frequency and nominal power of the $I_\mathrm{rf}$ were fixed between 6 and 10\,GHz
and 20\,dBm, respectively. The $I_\mathrm{rf}$ generates
damping-like ($H_\mathrm{DL}$) and field-like ($H_\mathrm{FL}$) effective fields, as well as an Oersted field $H_\mathrm{Oe}$. An in-plane magnetic field was applied
with an angle $\phi_\mathrm{H}$ of 30$^\circ$ and 210$^\circ$ from the current direction.
In addition, the $\phi_\mathrm{H}$-dependent ST-FMR measurements were also performed to extract spin current with out-of-plane spin polarization (see details in Note~2, Supporting Information)
When $f$ and $H$ satisfy the FMR condition, the $H_\mathrm{DL}$, $H_\mathrm{FL}$, and $H_\mathrm{Oe}$ drive a precession of the magnetization
in the FM layer, which induces an oscillation of the device
resistance through the anisotropic magnetoresistance (AMR).
The oscillating resistance mixes with the $I_\mathrm{rf}$, yielding a
direct current mixing voltage $V_\mathrm{mix}$, the latter was measured using
a bias tee and a nanovoltmeter (Figure~S5, Supporting Information). All ST-FMR measurements were performed at room temperature.\\

\begin{figure}
		\centering
  \includegraphics[width=0.95\linewidth]{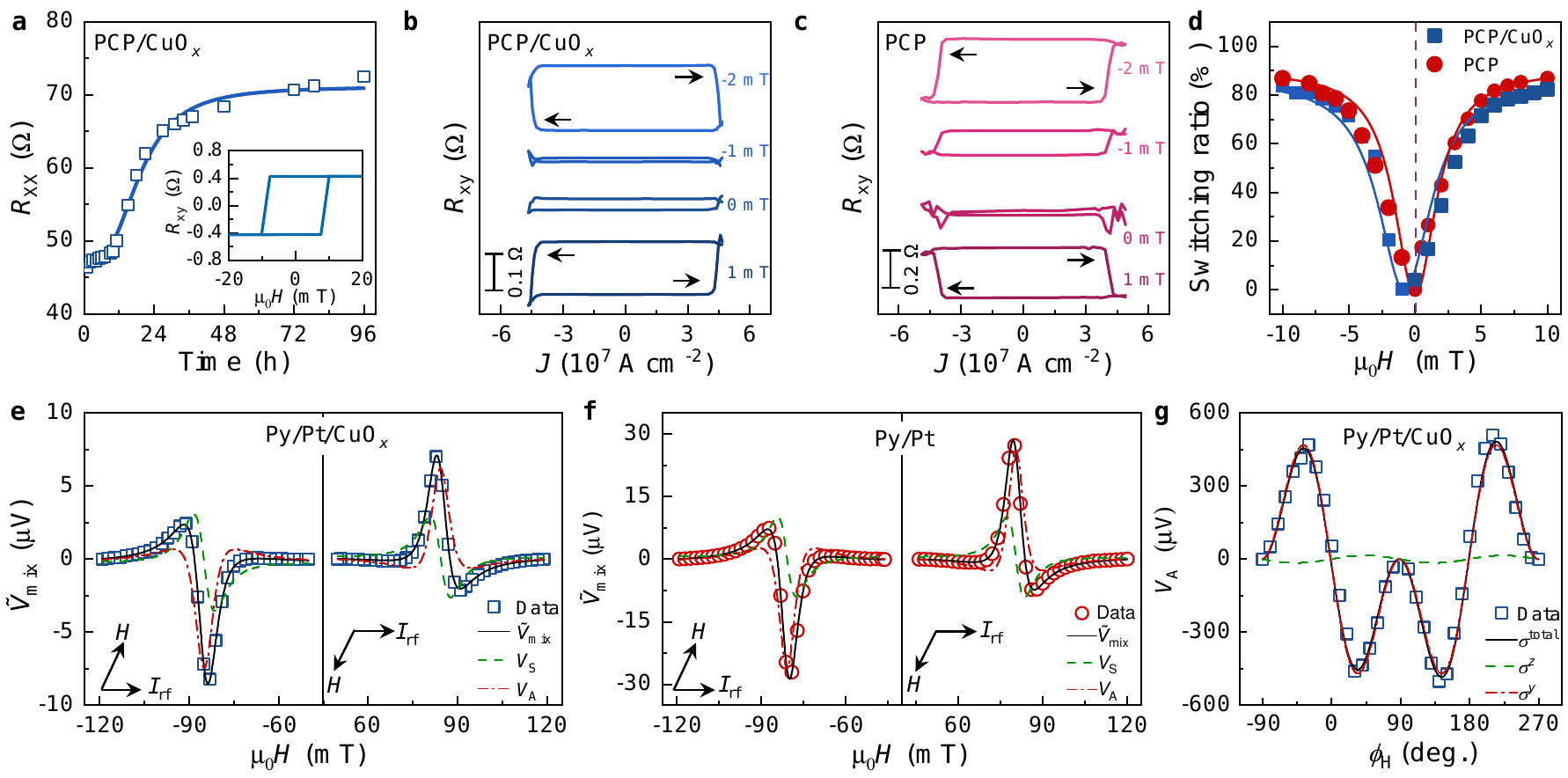}
  \caption{\label{fig:Nat.Oxid} Current-induced magnetization switching and ST-FMR.
  	 a) Evolution of longitudinal electrical resistance  $R_\mathrm{xx}$ with natural oxidation time for the PCP/CuO$_x$ heterostructures. The resistance increases with the natural oxidation time and starts to saturate after 24 hours of exposure in air.
  	The inset shows the field-dependent anomalous Hall resistance $R_\mathrm{xy}(H)$ after 24 hours of natural oxidation. For the $R_\mathrm{xy}$ measurements, the magnetic field was applied perpendicular to the film plane (i.e., $H_z$).
  	 b,c) Current-induced magnetization switching loops measured under various in-plane magnetic fields $H_x$ applied along the $x$-axis for PCP/CuO$_x$ (b) and PCP (c) heterostructures. The clockwise and anticlockwise $R_\mathrm{xy}$--$J$ loops were obtained by applying magnetic fields antiparallel (-$H_x$) or parallel ($H_x$) to the current direction, respectively.
  	Field and current directions are indicated in Figure~S2 in the Supporting Information.      
  	\tcr{For the PCP/CuO$_x$ heterostructures, the asymmetric magnetization switching loops for the positive and negative in-plane fields are due to the presence of out-of-plane spin polarization, which is induced by the oxidation gradient in the CuO$_x$ layer.}
  	 d) Magnetization switching ratio as a function of $H_x$ for PCP/CuO$_x$ and PCP heterostructures.  
    The magnetization switching ratio is defined as the ratio between the saturated Hall resistance in the $R_\mathrm{xy}$-$J$ and $R_\mathrm{xy}$-$H$ loops. The ratio saturates to $\sim$80\% when $H_x$ exceeds 7 mT for both heterostructures (see Figure~S4, Supporting Information). Note that field-free magnetization switching was observed in PCP/CuO$_x$, while it is absent in the PCP heterostructure.
  	e,f) ST-FMR spectra measured at 7\,GHz for Py/Pt/CuO$_x$ (e) and Py/Pt (f) heterostructures.
  	The Py (i.e., Ni$_{81}$Fe$_{19}$) layer offers an in-plane magnetic anisotropy for such measurements.	
  	Symbols are experimental data; solid lines represent fits to Equation~\eqref{eq:ST-FMR}. The dashed- and dash-dotted lines represent the symmetric ($V_\mathrm{S}$) and antisymmetric ($V_\mathrm{A}$) components of the resonance amplitude, respectively. 
  	The torque efficiency $\xi_\mathrm{FMR}$ can be obtained from the ratio between the magnitude of the symmetric $V_\mathrm{S}$ and antisymmetric component $V_\mathrm{A}$ of the spectra (see Equation~\eqref{eq:efficiency} in the Experimental Section). 
  	For these ST-FMR measurements, as indicated by the arrows in the insets, the angle $\phi_\mathrm{H}$ between the field $H$ and current $I_\mathrm{rf}$ directions was fixed at 30$^\circ$ and 210$^\circ$, respectively. 
  	By reversing the external magnetic field direction, the sign of $\tilde{V}_{\mathrm{mix}}$ also changes, as expected for the voltage generated by the ST-FMR.
  	g) Angular dependence of the antisymmetric resonance amplitude $V_\mathrm{A}$ for the Py/Pt/CuO$_x$ heterostructure. 
  	The black line ($\sigma^\mathrm{total}$) is a fit to Equation~S1 in the Supporting Information, while red and green lines represent the contributions with spin polarization along the $y$- ($\sigma^y$) and $z$-axis ($\sigma^z$), respectively. Note that the contribution with $\sigma^x$ (i.e.,  spin polarization along current direction) is negligible.  
  }
\end{figure}
\begin{figure}
		\centering
	\includegraphics[width=0.95\linewidth]{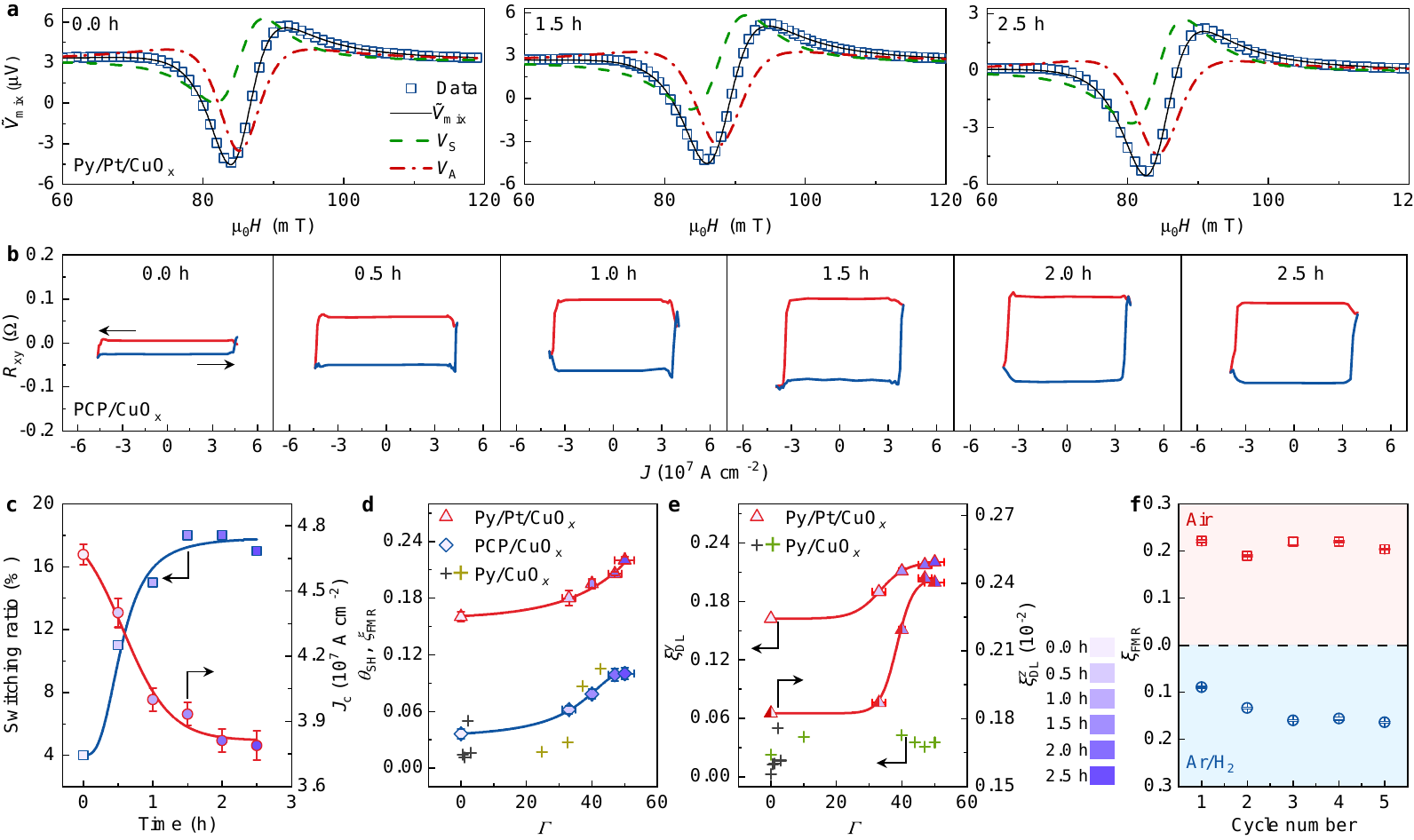}
	\caption{\label{fig:orbit_torque} Spin-orbit torque efficiency and critical current for magnetization switching.
		a) Representative ST-FMR spectra for Py/Pt/CuO$_x$ heterostructures annealed at 373\,K in the air atmosphere for 0, 1.5, and 2.5 hours, respectively. Solid lines are fits to Equation~\eqref{eq:First-order derivative of ST-FMR}, while dashed- and dash-dotted lines represent the symmetric and antisymmetric components of the corresponding spectra. 
		b) Field-free current-induced magnetization switching for PCP/CuO$_x$ heterostructures annealed at 373\,K in the air atmosphere for different times up to 2.5 hours. 
		\tcr{The results obtained by applying an in-plane magnetic field of 10 mT are summarized in Fig. S14 in the Supporting Information.}
	c) The magnetization switching ratio (left axis) and the critical electric current density $J_c$ (right axis) versus the annealing time for PCP/CuO$_x$ heterostructures.
		d) Spin-Hall angle $\theta_\mathrm{SH}$ and SOT efficiency $\xi_\mathrm{FMR}$ as a function of peak-area ratio $\Gamma$ (see definition in Figure~\ref{fig:XPS}). $\theta_\mathrm{SH}$ was obtained by harmonic-Hall measurements on  PCP/CuO$_x$ heterostructures annealed at 373\,K for different times, while the $\xi_\mathrm{FMR}$ was derived from ST-FMR measurements on Py/Pt/CuO$_x$ heterostructures. 
		e) In-plane ($\xi_\mathrm{DL}^y$, left axis) and out-of-plane ($\xi_\mathrm{DL}^z$, right axis) damping-like SOT efficiencies  
		versus $\Gamma$ for PCP/CuO$_x$ heterostructures annealed at 373\,K for different times. Note that $\xi_\mathrm{DL}^z$ is smaller than $\xi_\mathrm{DL}^y$ by almost a factor of 100.
		Both $\xi_\mathrm{DL}^y$ and $\xi_\mathrm{DL}^z$ were 
		determined through the angular-dependent ST-FMR spectra (see Figure~S18 and Note~2, Supporting Information).
		The results of Py/CuO$_x$ in panels (d) and (e) were taken from Refs.~\cite{an_electrical_2023,kageyama_spin-orbit_2019,gao_intrinsic_2018}.
		The torque efficiencies of Py/CuO$_x$ heterostructures are relatively small due to the lack of SOC layer, which is crucial to convert the orbital current into spin current (Figure~\ref{fig:DFT}a). 
		f) Torque "switch". SOT efficiency $\xi_\mathrm{FMR}$ for Py/Pt/CuO$_x$ heterostructures annealed alternatively in the air (2.5 hours) and  Ar(95\%)/H$_2$(5\%) (80 hours) atmospheres at 373\,K. The oxidation states in the CuO$_x$ layer can be controlled by the redox cycle, and the cycle number indicates the sequence of annealing treatments.   
		The ST-FMR spectra and magnetic properties of these heterostructures are presented in Figure~S20 and Table S4 in the Supporting Information.}
\end{figure}

\noindent\threesubsection{Analysis of ST-FMR}\\
The measured ST-FMR signals $V_\mathrm{mix}$ can be expressed as the sum of symmetric
and antisymmetric Lorentzian functions:
\begin{equation}
	\label{eq:ST-FMR}
	V_\mathrm{mix} = V_\mathrm{S}\frac{W^2}{(H - H_\mathrm{FMR})^2 + W^2} + V_\mathrm{A}\frac{W(H - H_\mathrm{FMR})}{(H - H_\mathrm{FMR})^2 + W^2}, 
\end{equation}
where $V_\mathrm{S}$ and $V_\mathrm{A}$ are the magnitude of the symmetric and antisymmetric component of the spectra; $W$ and $H_\mathrm{FMR}$ are the linewidth and the FMR resonance field, respectively. Generally, $V_\mathrm{S}$ arises solely from the damping-like torque $\left[\tau_{\mathrm{DL}} \propto \boldsymbol{m} \times(\boldsymbol{m} \times \boldsymbol{y})\right]$, while  $V_\mathrm{A}$ originates jointly from the field-like torque $\left(\tau_{\mathrm{FL}} \propto \boldsymbol{m} \times \boldsymbol{y}\right)$ and current-induced Oersted field,
respectively. Then, 
the charge to spin conversion efficiency $\xi_\mathrm{FMR}$ can be determined by the ratio of $V_\mathrm{S}$/$V_\mathrm{A}$ following the equation:
\begin{equation}
	\label{eq:efficiency}
	\xi_\mathrm{FMR} = \frac{2e}{\hbar}\frac{J_\mathrm{s}}{J_\mathrm{e}} = \frac{V_\mathrm{S}}{V_\mathrm{A}}\frac{e\mu_0M_\mathrm{s}t_\mathrm{NM}t_\mathrm{FM}}{\hbar}\sqrt{1+ \frac{M_\mathrm{eff}}{H_\mathrm{FMR}}},
\end{equation}
where $t_\mathrm{NM}$ and $t_\mathrm{FM}$ are the thickness of NM and FM layers, respectively; $H_\mathrm{FMR}$ is the resonance field in the FMR spectra; $M_\mathrm{s}$ and $M_\mathrm{eff}$ are the saturation and effective saturation magnetization, the latter can be extracted by the Kittel equation $\frac{2\pi f}{\gamma}$ = $\sqrt{H_\mathrm{FMR}(H_\mathrm{FMR} + M_\mathrm{eff})}$, with $\gamma$ being the gyromagnetic ratio. It is noted that in Figures~\ref{fig:Nat.Oxid} and~\ref{fig:orbit_torque}, the first derivatives of the mixed voltage $V_{\text{mix}}$ with respect to the magnetic field $H$ were presented:

\begin{equation}
	\begin{aligned}
		\label{eq:First-order derivative of ST-FMR}
		&\tilde{V}_{\mathrm{mix}} =\frac{\partial V_{\mathrm{mix}}}{\partial H} = \\
		& \frac{ -2V_\mathrm{S} W^2 (H-H_{\mathrm{FMR}}) + V_\mathrm{A} W \left[ W^2 - (H-H_{\mathrm{FMR}})^2 \right] }{ \left[ (H-H_{\mathrm{FMR}})^2 + W^2 \right]^2 }
	\end{aligned}
\end{equation}\\

\noindent\threesubsection{Harmonic Hall measurements}\\
For the ac harmonic-Hall measurements, a sinusoidal voltage with constant amplitude $V(t)$ =  $V_0$cos(2$\pi$$ft$) was applied using a lock-in
amplifier (Stanford Research Systems) to the Hall bar device with a reference frequency of 13.7~Hz. Two other lock-in amplifiers (signal recovery SR830) were employed to simultaneously measure the in-phase first and out-of-phase second harmonic voltage, while sweeping the applied magnetic field in different orthogonal directions.
Quantitatively, the damping-like and field-like fields are determined by the following expressions:
\begin{equation}
	\label{eq:harmonic}
	\Delta H_{x(y)}= \left(\frac{\partial V_{2 \omega}}{\partial H_{x(y)}}\right) / \frac{\partial^2 V_\omega}{\partial H_{x(y)}^2},
\end{equation}
\begin{equation}
	\label{eq:harmonic-DL}
	H_\mathrm{DL}/J=-2 \frac{\Delta H_x \pm 2\xi \Delta H_y}{J(1 - 4\xi^2)},
\end{equation}
\begin{equation}
	\label{eq:harmonic-FL}
	H_\mathrm{FL}/J = -2 \frac{\Delta H_y \pm 2\xi \Delta H_x}{J(1 - 4\xi^2)},
\end{equation}
where $\xi = R_\mathrm{PHR}/R_\mathrm{AHR}$ is the ratio between planar Hall resistance $R_\mathrm{PHR}$ and anomalous Hall resistance $R_\mathrm{AHR}$, and $\pm$ denotes +$M_z$ and -$M_z$ magnetic states (see details in Figure~S17, Supporting Information). Then, the torque efficiency or spin-Hall angle can be calculated according to 
\begin{equation}
	\label{eq:spin-hall-angle}
	\theta_{\mathrm{SH}} = 2eM_\mathrm{s}t_\mathrm{FM}H_\mathrm{DL}/\hbar J, 
\end{equation}
where $e$ is the electron charge, $t_\mathrm{FM}$ the thickness of the FM layer, $\hbar$ the reduced Planck constant, and $J$ the charge current density.\\
 
\noindent\threesubsection{Microstructure characterization}\\
Cross-sectional TEM heterostructures were prepared using a dual-beam focused
ion beam system (Helios G4 UX, Thermo Fisher). The W film was coated on
the sample surfaces to protect them from Ga ion damage. The
samples were attached to TEM grids and thinned to a medium
thickness before final grinding, followed by ion milling at 2 kV to
remove the amorphous layer to achieve electron beam transparency. Atomic resolution HAADF-STEM measurements were
performed using a 300 kV spherical aberration (Cs)-corrected
STEM (AC-STEM, JEM-ARM300F, JEOL). The EELS measurements were performed to check the distributions of various elements at different depths from the surface to the subsurface. \\ 

\noindent\threesubsection{First-principles calculations}\\
We constructed CuO$_x$/Cu heterostructures, which consist of four layers of metallic Cu and CuO$_x$ oxides. The O atoms are in contact with the metallic Cu atoms at the interfaces. The metallic Cu layer adopts a Cu(111) configuration, and 
the lattice parameters of CuO$_x$ layer used for the calculation are listed in Table~S2 in the Supporting Information.
First-principles calculations were performed based on the density functional theory (DFT), as implemented in the Vienna ab-initio Simulation Package (VASP), and the projector augmented wave (PAW) pseudo potentials were adopted in the calculations. We treated the exchange-correlation potential based on the Perdew-Burke-Ernzerhof generalized gradient approximation (PBE-GGA) functional, and a Hubbard U ($\sim$5\,eV) correction (GGA+U) was introduced to account for the strong correlation of Cu 3$d$ electrons. A 20~$\AA$ vacuum layer was included to eliminate interactions between periodic images. A plane-wave basis set with a kinetic energy cutoff of 500~eV was employed, and the Brillouin zone was sampled using a $\Gamma$-centered 8 $\times$ 8 $\times$ 1 Monkhorst-Pack $k$-point grid. Structural optimization was performed until the energy and force converge to 1 $\times$ 10$^{-5}$~eV and 0.01~eV/$\AA$, respectively.
\tcr{For the calculations, instead of polycrystalline CuO$_x$, the CuO$_x$/Cu heterostructures were built, and fully relax the atomic structures, which allows us to release the interfacial stress while studying the influence of different hybridization between CuO$_x$ and metallic Cu.}\\



\medskip
\noindent\textbf{Conflict of Interest}\\
The authors declare no conflict of interest.

\medskip
\noindent\textbf{Data Availability Statement} \\
The data that support the findings of this study are available from the corresponding author upon reasonable request.

\medskip
\noindent\textbf{Supporting Information}\\
Supporting Information is available from the Wiley Online Library or from the author.

\medskip
\noindent\textbf{Acknowledgements}\\
We thank Jianzhou Zhao for fruitful discussions.
This work was supported by the National Natural Science Foundation of China (Grant Nos. 12374105, 62274066, 12350710785, and 12561160109),
and the Fundamental Research Funds for the Central Universities. \\
\medskip
%

\bibliographystyle{MSP}

\bibliography{CuOx.bib}
\clearpage

\clearpage

\clearpage

\clearpage

\clearpage

\clearpage






\end{document}